\documentclass[conference]{IEEEtran}
\IEEEoverridecommandlockouts
\usepackage{cite}
\usepackage{amsmath,amssymb,amsfonts}
\usepackage{algorithmic}
\usepackage{graphicx}
\usepackage{textcomp}
\usepackage{xcolor}
\def\BibTeX{{\rm B\kern-.05em{\sc i\kern-.025em b}\kern-.08em
    T\kern-.1667em\lower.7ex\hbox{E}\kern-.125emX}}

\usepackage{hyperref}
\usepackage{listings}
\begin{document}

\title{Qmod: Expressive High-Level Quantum Modeling}

\author{
\IEEEauthorblockN{
    Matan Vax,
    Peleg Emanuel,
    Eyal Cornfeld,
    Israel Reichental,
    Ori Opher,
    Ori Roth,\\
    Tal Michaeli,
    Lior Preminger,
    Lior Gazit,
    Amir Naveh,
    Yehuda Naveh
}
\IEEEauthorblockA{
    \textit{Classiq Technologies}\\
    3 Daniel Frisch Street, Tel Aviv-Yafo, 6473104, Israel \\
    Emails: \{matan, peleg, eyal, israel, ori, oriroth, talmi, liorp, lior, amir, yehuda\}@classiq.io
}
}
\maketitle

\begin{abstract}
Quantum computing hardware is advancing at a rapid pace, yet the lack of high-level programming abstractions remains a serious bottleneck in the development of new applications. Widely used frameworks still rely on gate-level circuit descriptions, causing the algorithm’s functional intent to become lost in low-level implementation details, and hindering flexibility and reuse. While various high-level quantum programming languages have emerged in recent years -- offering a significant step toward higher abstraction -- many still lack support for classical-like expression syntax, and native constructs for useful quantum algorithmic idioms. This paper presents \emph{Qmod}, a high-level quantum programming language designed to capture algorithmic intent in natural terms while delegating implementation decisions to automation. Qmod introduces quantum numeric variables and expressions, including digital fixed-point arithmetic tuned for compact representations and optimal resource usage. Beyond digital encoding, Qmod also supports non-digital expression modes -- phase and amplitude encoding -- frequently exploited by quantum algorithms to achieve computational advantages. We describe the language’s constructs, demonstrate practical usage examples, and outline future work on evaluating Qmod across a broader set of use cases.
\end{abstract}


\bstctlcite{IEEEexample:BSTcontrol}

\section{Introduction}
Recent years have seen high-paced advances in quantum hardware. In anticipation of more reliable and scalable hardware, the development of practical applications is increasingly perceived as a bottleneck for the commercial utility of quantum technology. A growing number of researchers and engineers are participating in this effort, but the lack of strong abstractions and scalable tools slows their progress \cite{murillo2024}.

Quantum circuits are still the primary model for programming gate-based quantum computers today. The most widely used frameworks expose circuit terms – gates applied to specific qubits – to algorithm designers \cite{Qiskit2024, Pennylane}. However, describing non-trivial quantum logic at the circuit level can be extremely difficult. In particular, it is hard to do the following:
\begin{itemize}
\item Capture functionality in a way that communicates the intent clearly, abstracting from implementation decisions and details.
\item Express algorithms in their full generality, without hard-coding parameters of a specific application context.
\item Decompose descriptions into reusable building blocks.
\item Implement functionality efficiently in terms of hardware resource requirements.
\item Adapt and optimize implementations for different hardware and simulation platforms with their constraints and restrictions.
\end{itemize}
The exploration and development of quantum applications calls for a higher level of abstraction in formal descriptions and more powerful tools to process them \cite{dimatteo2024}.

The present state of quantum computing is comparable to the early days of classical computing, before high-level programming languages or hardware description languages had become widespread. Advances in languages and tools in classical hardware and software design have revolutionized their respective fields. Drawing analogies from these developments offers valuable insight into the path forward for quantum languages. Language concepts, such as functions, variables, types, expressions, and control-flow statements, can be brought into the quantum domain, essentially retaining their meaning and use. Conserving common linguistic intuitions, conventions, and terminology eases adoption and increases productivity.

At the same time, the unique characteristics of quantum computing call for special treatment in the design of a programming language. First, operations in quantum computation are generally reversible, with measurement being an exception. In addition, quantum states cannot be cloned \cite{nielsen2010quantum}. These restrictions can be integrated into the semantics of the language \cite{Qsharp2018}. Another concern is the disposal of intermediate computation values and the recycling of storage. Unlike the case with classical computing, reusing quantum storage requires explicit and often costly un-computation of temporary values to disentangle auxiliary qubits and return them to their initial zero state \cite{Silq2020}.

A key consideration in designing a quantum programming language is recognizing that the scarcity of resources -- such as qubits, gates, and coherence time -- fundamentally alters the compilation problem compared to classical software or hardware. This seems to be the case in the foreseeable future, even as hardware technology evolves and matures. One implication is the need for compact representation of numbers. In classical programming, bit-level control over the encoding of integers and fraction numbers is rarely required. Compilers typically rely on hardware-level support for storing and manipulating numbers in 64-bit registers. In contrast, quantum hardware is far from accommodating such extensive storage requirements, and quantum algorithms are expected to be useful with much less. Hence, bit-level flexibility is needed even with high-level quantum languages. A compiler should be able to tune the circuit-level implementation of numeric operations to the domains of variables and the required precision on a per-case basis.

In quantum algorithms, qubits often encode algorithmic-level data digitally in the computational basis. Arithmetic operations on digitally encoded numbers can be implemented using logic analogous to classical bits, albeit reversible. However, in some situations, it is preferable to compute arithmetic operations in other ways, for example, in the Fourier basis \cite{draper2000}. Moreover, quantum states often encode information non-digitally, for example, in the amplitudes or the phases of computational-basis states. Transitions between these modes of encodings are a recurring pattern in quantum algorithms \cite{nielsen2010quantum}. We argue that language support for amplitude and phase encoding of numeric values can dramatically improve the conciseness and readability of the code in these cases. More importantly, synthesizing the gate-level implementation to perform computations directly in these modes can significantly reduce circuit depth and other computational resources.

In this paper we present Qmod -- a language for describing quantum algorithms in terms that express high-level functional or algorithmic intent, leaving out implementation details and paving the way to strong automation \cite{classiq2024, Qmod_docs}. Qmod draws on previous work in high-level quantum programming languages \cite{Quipper2013, Qsharp2018, Silq2020}. However, Qmod goes beyond existing languages in key capabilities, and specifically in supporting quantum numeric variables and expressions. With these constructs, Qmod addresses the challenges and considerations raised above. The rest of this paper is structured as follows: Section 2 is a general language walkthrough, Section 3 dives into the constructs for arithmetic expressions in different encoding modes, and Section 4 demonstrates the application and utility of these constructs through two real-life usage examples. Section 5 discusses related work, and Section 6 concludes.

\section{Qmod fundamentals}
\subsection{Input formats}
Qmod is designed as a domain-specific language with two alternative input formats: an external domain-specific language (DSL) with its dedicated syntax, and an embedded DSL with Python as host language. These two formats are equivalent with respect to the set of constructs that they expose and are automatically interconvertible. Each has its advantages, and programmers can code in their preferred way. In this paper, we shall use the Qmod flavor that is embedded in Python. The reason for this choice is the flexibility of compile-time computation enabled by Python as the host language.

Qmod's Python embedding utilizes Python's language features to encode domain-specific constructs in a manner consistent with standard Python practices. This includes type-hint annotations, decorators, special methods for operator overloading, and lambda expressions for managing lexical scopes. When the Python code is executed using a regular Python interpreter, it generates a representation of the Qmod description, which is then compiled using specialized tools.

\subsection{Quantum objects variables and functions}
The overall state of a quantum computer with \(n\) qubits is a vector in the \(2^n\)-dimensional Hilbert space, where the dimensions correspond to the \(2^n\) computational-basis states. However, from the high-level description perspective, the computer's state at a given point in the computation represents data items that are operated on by the computation, in particular numeric values. We view these data items as objects, analogous to objects in classical programming languages. Quantum objects are allocated from the available quantum storage pool, their state evolves throughout their lifetime, and they are often released back so that the storage may be reused.

Quantum variables and quantum functions are two fundamental concepts in the Qmod language, with a clear analogy to their counterparts in classical programming languages. Quantum variables reference quantum objects, and quantum functions operate on quantum objects.

Quantum variables are declared either locally in a function scope or as function parameters. They need to be initialized explicitly, setting their reference to some quantum object. This is akin to the semantics of nullable variables in classical languages. Two basic ways to initialize a quantum variable are through the direct allocation of storage (qubits in the zero states) and as the left-value of an assignment statement, with the right-value being a numeric value or expression. Variables can also be initialized by passing them as output parameters of functions. The no-cloning rule is built into the semantics of quantum variables, by guaranteeing single-reference within a given scope. Specifically, assignment is performed "by value", copying computational basis state of the right-value expression to a zero-initialized left-value variable through entanglement. 

Quantum functions are the basic encapsulation unit in Qmod. Just like in classical programming languages, decomposing algorithms into functions, or subroutines, is a key to abstraction and reuse. Quantum functions may declare parameters of different types, in any of the three categories – quantum types, classical types, and function types. The functional hierarchy of a description serves as the starting point for the compiler-optimizer in breaking the optimization problem into more loosely coupled sub-problems.

An example of a Qmod description to prepare and measure the bell state is shown in \autoref{fig:bell_state_example}. In the Python embedding of Qmod, quantum functions are designated with the decorator \lstinline{@qfunc}, and their parameters must be declared with type-hint annotations. Function \lstinline{main} is the entry point of the quantum program, and its output parameters are measured and returned to the caller. In this example, \lstinline{main} allocates the 2-qubit array \lstinline{res}, passes it to function \lstinline{bell}, and outputs it for measurement. Note that both parameter \lstinline{res} of function \lstinline{main} and parameter \lstinline{pair} of function \lstinline{bell} are declared as qubit arrays of size 2. But \lstinline{res} unlike \lstinline{pair} is modified with the generic class Output, which indicates that it is an output-only parameter. Therefore, variable \lstinline{res} must be initialized prior to its use in \lstinline{main}.  Note that the \textit{import} statement is shown in this listing, but omitted in subsequent examples.

\begin{figure}[ht]
\begin{lstlisting}
from classiq.qmod import *

@qfunc
def bell(pair: QArray[QBit, 2]):
  H(pair[0])
  CX(pair[0], pair[1])


@qfunc
def main(res: Output[QArray[QBit, 2]]):
  allocate(res)
  bell(res)
\end{lstlisting}
\caption{
Code listing for the bell state preparation. When this code executes \lstinline{res} is sampled with value [0, 0] or [1, 1] in equal probability.
}
\label{fig:bell_state_example}
\end{figure}

\subsection{Quantum types}
Quantum variables have fixed static types -- either scalars, arrays, or structs. Scalars generally represent numbers. Single bit variables, declared with the class \lstinline{QBit}, represent the values 0 and 1, or the Boolean values \lstinline{True} and \lstinline{False}. Numeric variables, declared using the generic class \lstinline{QNum}, represent numbers in some discrete domain – integer or fixed-point real. Variables of all types expose the attribute \textit{size} to query the overall number of qubits used to store the respective quantum object. Two additional attributes are associated with \lstinline{QNum} variables, determining their numeric interpretation: the number of fractional digits and the use of a sign bit to represent signed numbers. These properties are either specified explicitly in the variable declaration or inferred upon its initialization, and are subsequently immutable. Numeric type inference is discussed in Section 3.1.

Scalars can be aggregated into composite data structures using array and structure types. A quantum array, declared with the generic class \lstinline{QArray}, is an object that supports indexed access to parts of its state -- its elements -- using the subscript operator (square bracket). A quantum structure, defined as a Python subclass of \lstinline{QStruct}, supports named access to parts of its state -- its fields -- using the dot operator. Thus, parts of a composite structure can be accessed using conventional path expressions.

The example in \autoref{fig:quantum_struct_example} demonstrates the use of aggregate types in Qmod. \lstinline{MyStruct} declares field \lstinline{data} of a numeric array type, and a numeric field called \lstinline{sum}. Function \lstinline{main} allocates an instance of the struct, and applies the Hadamard transform to field \lstinline{data}, thus putting all its elements in an equal superposition of the values in their domain. Subsequently, a \textit{repeat} statement iterates over these elements and integrates their values into field \lstinline{sum} (see more on in-place addition in Section 3.1). Note that in passing array \lstinline{s.data} to function \lstinline{hadamard_transform}, its type is converted to a qubit array view per the type of parameter \lstinline{qba}. This auto-conversion between the structured and packed view of quantum objects is key in defining reusable building blocks. In this case, function \lstinline{hadamard_transform} is defined in a generic way and can be reused in other contexts. 

\begin{figure}[ht]
\begin{lstlisting}
class MyStruct(QStruct):
  data: QArray[QNum[2], 3]
  sum: QNum[4]

@qfunc
def hadamard_transform(qba: QArray[QBit]):
   repeat(qba.size, lambda i: H(qba[i]))

@qfunc
def main(s: Output[MyStruct]):
  allocate(s)
  hadamard_transform(s.data)
  repeat(
    s.data.len,
    lambda i: inplace_add(s.data[i], s.sum)
  )
\end{lstlisting}
\caption{
An example of the use of a quantum struct type. When this code executes \lstinline{s}  is sampled with field \lstinline{sum} corresponding to the sum of the elements in \lstinline{data}.
}
\label{fig:quantum_struct_example}
\end{figure}

The example in \autoref{fig:quantum_struct_example} also shows the syntax of the Qmod iterative statement \textit{repeat} as it is coded in Python. It uses a Python function \lstinline{repeat}, taking the iteration count and the loop body as arguments. Note that the loop count in these cases is a Qmod expression, which is represented symbolically in Python, so a regular Python \textit{for} statement cannot be used. In cases where the count value is known, a \textit{for} loop can be used to expand the same Qmod statements multiple times. This is functionally equivalent to using \textit{repeat}, but is often less efficient than leaving the loop rolled up for the Qmod compiler to process. The loop body is passed as a lambda expression taking the iteration variable as parameter. Similar syntactic structure is used for the Qmod \textit{if} statement, and for the quantum functors: \textit{control}, \textit{invert}, and \textit{power}.

\subsection{Uncomputation and release}
The conjugation pattern \(U^{\dagger} V U\) is ubiquitous in quantum computation. Mathematically, it can be viewed as applying the operation \(V\) in the basis defined by the operation \(U\). In practice, this construction also provides a structured way to prepare and then uncompute intermediate results, resetting auxiliary qubits that are used to store them.

In Qmod, the \textit{within-apply} statement expresses the conjugation \(U^{\dagger} V U\), where \(U\) and \(V\) correspond to the outer and inner statement blocks respectively. To the extent that the outer block does not introduce superposition into the quantum objects it operates on, and the inner block does not modify the state of these objects, they subsequently return to their state prior to the \textit{within-apply} statement. Specifically, quantum objects that are allocated under the outer block, or in functions called by it, are uncomputed back to the zero state, and returned to the pool of clean qubits for subsequent reuse. Variables that are initialized in the outer block and are still in scope go back to be uninitialized (“null reference”).

In \autoref{fig:phase_flip_example} function \lstinline{flip_phase} takes a quantum function \lstinline{predicate} as parameter, and applies a \(\pi\) phase shift to states marked by it. Variable \lstinline{aux} is initialized inside the \textit{within} block, and is used as the result qubit in \lstinline{predicate}. The inversion of the \textit{within} block, which follows the \textit{apply} block, returns the qubit to the zero state. Subsequently, \lstinline{aux} is left uninitialized, and the qubit is released to be recycled in later operations. Note that the Python generic class \lstinline{QCallable} is used to designate a function parmater in Qmod, with the expected arguments listed as generic parameters. The argument for \lstinline{predicate} can be a named quantum function, or a lambda (anonymous) function specified inline, optionally with variables captured from the enclosing scope.

Function \lstinline{main} in this listing demonstrates the use of function \lstinline{flip_phase}. The argument for \lstinline{predicate} is a lambda function that captures variable \lstinline{qba} and simply marks its state \(\left|11\right\rangle\). The state vector obtained when simulating this description shows equal magnitudes of all 4 amplitudes of \lstinline{qba} with \(\pi\) phase on the state \(\left|11\right\rangle\) relative to the other states.

\begin{figure}[ht]
\begin{lstlisting}
@qfunc
def flip_phase(predicate: QCallable[QBit]):
  aux = QBit()
  within_apply(
    within=lambda: [
      allocate(aux),
      predicate(aux),
    ],
    apply=lambda: Z(aux),
  )

@qfunc
def main(qba: Output[QArray[QBit, 2]]):
  allocate(qba)
  hadamard_transform(qba)
  flip_phase(lambda target: CCX(qba, target))
\end{lstlisting}
\caption{
An example of the use of \textit{within-apply} statement to allocate and release a qubit, as well as the use of a lambda function with captured variables as a parameter to another user-defined quantum function.
}
\label{fig:phase_flip_example}
\end{figure}

\section{Quantum arithmetic expressions}
Arithmetic computations make up a necessary part of numerous quantum algorithms, either directly or as means to approximate other mathematical functions \cite{häner2018}. Moreover, arithmetic expressions often provide natural terms for embedding problem-domain classical functions into quantum subroutines. Common examples are oracles in search problems and cost functions in optimization problems.

In Qmod, quantum numeric variables, along with classical variables and literal constants, can be composed into arbitrarily complex expressions. This is done much like in classical programming languages using native infix operators – arithmetic, bitwise, relational, and logical. For example, if \lstinline{x} and \lstinline{y} are quantum numeric variables, the expression \lstinline{2.5*x + y <= 10} is a legal quantum expression that evaluates to a Boolean value. To the extent that \lstinline{x} and \lstinline{y} are in a superposition state the expression evaluates to the corresponding superposition. Quantum expressions can occur as the right-value of digital assignment, as the Boolean condition of control statements – the quantum counterpart of classical \textit{if} statements, and in special arithmetic evaluation encodings.

\subsection{Digital arithmetic}
The operator for digital assignment of an arithmetic expression \(f\) over a set of quantum variables \(x_i\) (with an overall size of \(n\) qubits), given a target variable of size \(m\) is described mathematically thus –
\begin{equation}
\left|x_i\right\rangle_n\left|0\right\rangle_m\ \rightarrow  \left|x_i\right\rangle_n\left|f(x_i)\right\rangle_m
\end{equation}

Assignments in Qmod can take one of three forms: out-of-place assignment, in-place-xor assignment, and in-place-add assignment. The out-of-place assignment is the fundamental operation, while the other two forms can be defined in its terms. The out-of-place assignment requires an uninitialized numeric variable as its left-value (target) variable. The variable subsequently references a newly allocated quantum object, which stores the result. The new object is of type \lstinline{QNum} with an overall size in bits, signedness, and number of fractional digits, such that its domain tightly covers the range of possible expression values. The domain is inferred through the static analysis of the expression, given the types of quantum variables, classical constants, and how they are composed through operators. This is done using interval arithmetic, a common practice in abstract interpretation \cite{goubault2013}. Tight domain representation has a dramatic impact on the implementation's efficiency.
    
The code example in \autoref{fig:digital_arithmetic_example} demonstrates a simple digital arithmetic computation. An expression over quantum variables \lstinline{a} and \lstinline{b} is assigned to variable \lstinline{res}, which is subsequently measured. Variable \lstinline{a} is declared as a 2-qubit unsigned integer, representing values in the domain [0, 1, 2, 3]. Variable \lstinline{b} is declared as a 2-qubit signed number with one fractional digit, representing values in the domain [-1.0, -0.5, 0, 0.5]. Using fixed-point interval arithmetic, the range of expression values is statically determined to be covered by the domain [0, 0.125, … 1.875]. Thus the type of res is inferred at the point of assignment to be a 4-qubit unsigned number with 3 fractional digits. Specifically, the number of fractional digits, in this case, is the result of multiplying the constant 0.125 (2 fractional digits) with the integer a (0 fractional digits) and the fixed-point variable b (1 fractional digit). Note that out-of-place assignment in the Qmod Python embedding uses the operator \lstinline{|=} since Python does not allow overloading the operator \lstinline{=}. Assignment operators are also supported through function-call syntax, and specifically the function \lstinline{assign} in the case of out-of-place assignment.

\begin{figure}[ht]
\begin{lstlisting}
@qfunc
def main(res: Output[QNum]):
  a = QNum(size=2)
  a |= 3

  b = QNum(size=2, is_signed=True, fraction_digits=1)
  allocate(b)
  hadamard_transform(b)

  res |= 0.25*a*b + 1.5
\end{lstlisting}
\caption{
An example of digital arithmetic and numeric type inference. Executing this code samples \lstinline{res} in one the values: 0.75, 1.125, 1.5, or 1.875, with equal probability, for the \lstinline{b} values -1.0, -0.5, 0, 0.5, respectively.
}
\label{fig:digital_arithmetic_example}
\end{figure}

The two in-place forms require an initialized quantum variable as the left-value. The result of the right-value expression is evaluated in the same way as with out-of-place assignment, and is subsequently xor-ed with or added to the value of the left-value variable. The result is then stored back onto the same object without changing its size or numeric representation attributes. The compiler may need to apply truncation and sign-extensions to align the expression result with the target object. Qubits either truncated or used for sign extension are automatically uncomputed and released.
In \autoref{fig:inplace_xor_example} function \lstinline{my_oracle} marks states in which elements in a numeric array are in ascending order with a negative phase. It uses the function \lstinline{flip_phase} from \autoref{fig:phase_flip_example}, passing it a lambda function that XORs the expression value with \lstinline{res} using the \textit{inplace\_xor} statement. This function can be used, for example, as the oracle in the Grover search algorithm \cite{grover1996fast}, to find states that satisfy the predicate. Note that the bitwise-and operator (\lstinline{&}) is used here instead of the native Python logical \lstinline{and}, as the latter with its short-circuiting semantics cannot be overloaded.

\begin{figure}[ht]
\begin{lstlisting}
@qfunc
def my_phase_oracle(arr: QArray[QNum, 3]):
  flip_phase(lambda res:
    inplace_xor(
      (arr[0] < arr[1]) & (arr[1] < arr[2]),
      res,
    )
  )
\end{lstlisting}
\caption{
An example of the use of \textit{inplace\_xor} statement to implement a phase oracle for a sorted array.
}
\label{fig:inplace_xor_example}
\end{figure}

The number of bits required to store both the final and intermediate results of a composite expression can grow rapidly with fixed-point arithmetic. In particular, preserving full precision in multiplication means the number of fractional digits in the product is at least the sum of the fractional digits in the multiplicands. Moreover, classical numeric values (given as Python literals or compile-time variables) often use 64-bit floating-point representation, which can require an even larger number of bits to be represented exactly in fixed-point format. In \autoref{fig:digital_arithmetic_example}, for example, \lstinline{a} is an integer, \lstinline{b} has one fractional digit, and the constants \(0.25\) and \(1.5\) can be represented with two and one fractional digits, respectively. Hence, without loss of precision, the value of the expression \lstinline{0.25*a*b + 1.5} requires three fractional digits. If, instead of \(0.25\), we had \(0.3\) (decimal), preserving the exact floating-point value would require all mantissa bits (typically 53).

Full precision is practically not feasible, nor required, for the purpose of quantum algorithms. With arbitrarily composed expressions, arithmetic needs to be performed in some precision-bound context. In Qmod, a configurable “machine-precision” maximum is applied globally, across all arithmetic computations in a description. This bound can be further refined in specific contexts when the number of fractional digits of the target variable is known. Truncation of intermediate and final results based on required precision may need to take place, and the truncated qubits must be uncomputed and released.

The gate-level implementation that is synthesized to implement digital arithmetic expressions often requires auxiliary qubits to store intermediate results. Auxiliaries must be correctly uncomputed and released. The compiler can generate an optimal gate-level implementation tailored for the specific expression and the required circuit properties. Implementation trade-offs, in terms of depth, auxiliary requirements, C-NOT gate counts, etc., can be automatically handled by the compiler at the single operator level and across the full expression \cite{Meuli_Soeken_et_al2019}.

\subsection{Arithmetic in phase}
Quantum states can also encode data in the phase, that is, in the Z-axis angles of computational-basis states relative to each other. Phase-encoded data are not accessible for direct measurement, but quantum algorithms explicitly exploit it to gain an advantage. Examples of such strategies are phase flip in oracular algorithms and phase arithmetic in the Fourier basis. A highly general and useful set of functions can be computed directly in the phase of quantum states, namely quadradic binary functions, and functions reducible to them. This is due to the fact that the evolution of Ising-model Hamiltonians can be efficiently decomposed into gate-level circuits \cite{aharonov2005}.

Phase rotation for the arithmetic expression computing \(f\) over a set of quantum variables \(x_i\) (with an overall size of n qubits), given the coefficient \(\theta\), is described mathematically thus –
\begin{equation}
\left|x_i\right\rangle_n\rightarrow e^{i\theta f\left(x_i\right)}\left|x_i\right\rangle_n
\end{equation}

The corresponding operator is -

\begin{equation}
U_{\text{phase}}(f,\theta)
\;=\;
\begin{pmatrix}
e^{i \theta f(0)} &  &  \\
 & \ddots &  \\
 &  & e^{i \theta f\bigl(2^n - 1\bigr)}
\end{pmatrix}.
\end{equation}

The example in \autoref{fig:phase_arithmetic_example} demonstrates an arithmetic expression computed in the phase. In function \lstinline{main} the variable \lstinline{x} is put in a uniform superposition of computational-basis states, and subsequently the \textit{phase} statement is used to rotate the states at an angle of \(\frac{\pi}{4}x^2\). In the state vector obtained by executing this code in a simulator, state $\left|1\right\rangle$ has phase $\frac{\pi}{4}$ relative to state $\left|0\right\rangle$, state $\left|2\right\rangle$ has phase $\pi$, and state $\left|3\right\rangle$ again has phase $\frac{9\pi}{4}$ – a full $2\pi$ rotation plus $\frac{\pi}{4}$. Note that in this example, the application of \textit{phase} does not change the distribution of \lstinline{x} when sampling it.

\begin{figure}[ht]
\begin{lstlisting}
@qfunc
def main(x: Output[QNum]):
  allocate(x)
  hadamard_transform(x)
  phase(x**2 , pi/4)
\end{lstlisting}
\caption{
An example of performing arithmetic in the phase using the \textit{phase} statement. States of \lstinline{x} are rotated by \(\frac{\pi}{4}x^2\).
}
\label{fig:phase_arithmetic_example}
\end{figure}

\subsection{Arithmetic in amplitude}
In many quantum algorithms, the amplitudes of computational-basis states are used to encode and manipulate classical problem-domain data. Amplitude encoding has the inherent advantage of requiring logarithmic-size storage compared to the classical representation of the same data. Algorithms for matrix inversion and Monte Carlo integration are examples of the utilization of amplitude-encoded data \cite{HHL2009, montanaro2015}. In these cases, a key component of the algorithm is the transition from digitally encoded data to the amplitude-encoded result of some computation.

This transition requires a unitary operation that computes a function over quantum variables and encodes its result in the amplitude of the corresponding quantum state. Since the function itself is typically non-unitary, an additional "indicator" qubit (or qubits) must be used to preserve normalization - a technique generally known as block encoding. The result is represented in the amplitude of a designated state, while the remaining irrelevant amplitude is assigned to a state orthogonal to it. In the general case, such an operation can be implemented by applying uniformly controlled rotations on the indicator qubit, and there are methods for more compact implementations in important special cases \cite{mottonen2005}.

In Qmod this operation is natively supported using the amplitude-encoded arithmetic assignment statement. The amplitude encoding of an arithmetic expression computing \(f\) over a set of quantum variables \(x_i\) (with an overall size of \(n\) qubits) and an indicator qubit is mathematically described as follows -

\begin{equation}
\left|x_i\right\rangle_n\left|0\right\rangle\rightarrow\left|x_i\right\rangle_n(\sqrt{1-f\left(x_i\right)^2}\left|0\right\rangle+\ f(x_i)\left|1\right\rangle)
\end{equation}

Here, \(f\) is trimmed to take values in the range [-1, 1] for the domain of \(x_i\), per normalization requirement. The corresponding operator can be described thus -

\begin{equation}
U_{\text{amp}}(f)
\;=\;
\begin{pmatrix}
* & * \\
\begin{pmatrix}
f(0) &  & \\
 & \ddots & \\
 & & f(n-1)
\end{pmatrix} & * \\
\end{pmatrix}
\end{equation}

The lower left block is the diagonal computation of \(f\) values, while the purpose of the equally sized other blocks is to preserve unitarity of the overall matrix (* designates don't-care values).

\autoref{fig:amplitude_assignment_example} is a simple usage example of an arithmetic expression computed in the amplitude. In function \lstinline{tanh_amp} the value of an expression over \(x\) is stored in the \(\left|1\right\rangle\) amplitude of variable \lstinline{ind}. The expression is the first three non-zero terms in Taylor expansion the hyperbolic tangent around the value 0, hence \lstinline{tanh_amp} approximates this function in the amplitude of its output qubit. Function \lstinline{main} calls \lstinline{tanh_amp} and passes \(x\) initialized to the value 0.8125. Executing this function will sample the value 1 for \lstinline{ind} with probability close to \(\tanh(0.8125) \sim 0.671\).

\begin{figure}[ht]
\begin{lstlisting}
@qfunc
def tanh_amp(x: QNum, ind: Output[QBit]):
  allocate(1, ind)
  assign_amplitude(
    x - 1/3 * (x**3) + 2/15 * (x**5),
    ind,
  )

@qfunc
def main(
  x: Output[QNum[5, UNSIGNED, 5]],
  ind: Output[QNum],
):
  allocate(x)
  x ^= 0.8125
  tanh_amp(x, ind)
\end{lstlisting}
\caption{
An example of the use of \textit{assign\_amplitude} statement to approximate hyperbolic tangent of \lstinline{x} in the amplitude of the state in which \lstinline{ind} is 1.
}
\label{fig:amplitude_assignment_example}
\end{figure}

\section{Examples}
\subsection{Piecewise linear approximation of functions}

Arithmetic expressions can be used to compute useful functions via piecewise-polynomial approximation. We demonstrate here a special case, namely that of piecewise-\textit{linear} approximation. In Qmod we can directly express the linear function \(f_i(x) = a_ix+b_i\) for segment \(i\) as a quantum expression. We can conditionally apply the relevant function, depending on the segment within which \(x\) falls, using the \textit{control} statement. The coefficients \(a_i\) and \(b_i\) for each segment are computed classically using the Chebyshev interpolation.

The implementation of the quantum function \lstinline{compute_piecewise_linear}
is shown in \autoref{fig:piecewise_linear_example}. Given the classical arrays of coefficients \lstinline{a} and \lstinline{b} and the quantum numeric parameter \lstinline{x}, it computes the respective value \lstinline{f_x} in-place. First, the segment selector \lstinline{label} is computed from \lstinline{x}. For simplicity, we assume that the number of segments is a power of 2, and the partition is uniform. Thus, we only copy the most significant bits over to \lstinline{label}. Subsequently, we apply the respective linear function conditionally, based on \lstinline{label}.

\begin{figure}[ht]
\begin{lstlisting}
@qfunc
def init_label(
  v: QArray[QBit],
  label: Output[QArray[QBit]]
):
  allocate(int(log(NUM_SEGS, 2)), label)
  repeat(
    label.len,
    lambda i: CX(v[v.len - label.len + i], label[i])
  )

@qfunc
def compute_piecewise_linear(
  a: CArray[CReal], b: CArray[CReal],
  x: QNum, f_x: QNum,
):
  label = QNum()
  within_apply(
    lambda: init_label(x, label),
    lambda: repeat(a.len,
      lambda i: control(
        label == i,
        lambda: inplace_xor(a[i]*x + b[i], f_x),
      )
    )
  )
\end{lstlisting}
\caption{
Code listing of a function computing any piecewise-linear approximation.
}
\label{fig:piecewise_linear_example}
\end{figure}

The use of \lstinline{compute_piecewise_linear} to approximate the hyperbolic tangent in the range \([0,1)\) is shown in \autoref{fig:tanh_approximation_example}. Here we first compute the coefficients of the function \lstinline{np.tanh} for each segment classically using the Chebyshev interpolation (the implementation of \lstinline{linear_coefs} is not shown). We then allocate the result object and call \lstinline{compute_piecewise_linear}. Note that the precision of the result is controlled via the classical parameter \lstinline{p}, which is used to declare the number of fractional digits of \lstinline{f_x}. The arithmetic computation under \lstinline{compute_piecewise_linear} is automatically adapted to the numeric attributes of the target variable.

\begin{figure}[ht]
\begin{lstlisting}
@qfunc
def compute_tanh(
  p: CInt,
  x: QNum,
  f_x: Output[QNum["p", UNSIGNED, "p"]],
):
  a_coefs = []
  b_coefs = []
  for i in range(NUM_SEGS):
    a, b = linear_coefs(
      np.tanh, i/NUM_SEGS,
      (i+1)/NUM_SEGS,
    )
    a_coefs.append(a)
    b_coefs.append(b)

  allocate(f_x)
  compute_piecewise_linear(a_coefs, b_coefs, x, f_x)
\end{lstlisting}
\caption{
Code listing of a quantum function computing the piecewise-linear approximation of hyperbolic tangent function.
}
\label{fig:tanh_approximation_example}
\end{figure}

\subsection{Objectives and constraints in combinatorial optimization}

To demonstrate an application that combines two forms of arithmetic expression encoding, we turn to the integer knapsack problem, an example of a combinatorial optimization problem \cite{pferschy2004}. Simply put, the task is to determine how items of different types can be put in a knapsack to maximize their summed value, without exceeding a specified overall weight.
The mathematical formulation of the problem is as follows.

Given –

\(x_i\) – the number of items from each type (in the domain \(D_i\))

\(v_i\) – the value of each item

\(w_i\) – the weight of each item

find \(x\) that maximizes the value \(\Sigma_iv_ix_i\) constrained by the weight \(\Sigma_iw_ix_i\le C\).

For simplicity, we will look concretely at the following problem instance – we have \(a\) items of type \textit{A} where \(a\in\left[0..7\right]\), and \(b\) items of type \textit{B} where \(b\in\left[0..3\right]\). The values of the items are \(v_a=3\) and \(v_b=5\), the weights are \(w_a=2\) and \(w_b=3\), and the maximal weight is \(C=12\). Thus, our objective is to maximize \(3a+5b\), while satisfying the constraint \(2a+3b\le12\).

The approach we shall take is applying the Quantum Approximate Optimization Algorithm (QAOA) \cite{farhi2014quantum}. QAOA is a variational algorithm, where a parametric quantum circuit (ansatz) is used to span the space of possible solutions to a given optimization problem with a relatively small set of real-value parameters in a restricted domain. The algorithm explores this space, optimizing the parameters to minimize the estimated cost of the sampled result. Imposing problem constraints in QAOA is a challenging task, and several approaches have been suggested in recent years \cite{Blekos2024}. Here we present another valid method, combining two expression assignment forms in Qmod.

The ansatz in QAOA consists of an initialization layer, typically preparing a state of superposition of all possible solutions, followed by an alternating application of a cost operator and a mixer operator, each parameterized by a different rotation angle. The code for the quantum model is shown in \autoref{fig:knapsack_qaoa_full_example}. The details of the general QAOA algorithm are extensively discussed in the literature and will not be covered here. The definition of function \lstinline{cost_layer} is left out in this listing, and is discussed in detail below.

\begin{figure}[ht]
\begin{lstlisting}
class KnapsackVars(QStruct):
  a: QNum[3]
  b: QNum[2]

@qfunc
def init(v: QArray[QBit]):
  repeat(v.size, lambda i: H(v[i]))

@qfunc
def mixer_layer(v: QArray[QBit], beta: CReal):
  repeat(v.size, lambda i: RX(beta, v[i]))

@qfunc
def main(
  gammas: CArray[CReal, NUM_LAYERS],
  betas: CArray[CReal, NUM_LAYERS],
  v: Output[KnapsackVars],
):
  allocate(v)
  init(v)
  repeat(NUM_LAYERS, lambda i:[
    cost_layer(v, gammas[i]),
    mixer_layer(v, betas[i])
  )]
\end{lstlisting}
\caption{
Code listing of the QAOA ansatz for an integer knapsack problem.
}
\label{fig:knapsack_qaoa_full_example}
\end{figure}

First, we observe how quantum struct \lstinline{KnapsackVars} is used to aggregate the problem variables, in this case, two integers representing the number of \textit{A} items and \textit{B} items. The use of a struct lets us sample the output of \lstinline{main} and define the cost function in the problem terms, as we will see below. Parts of the algorithm that are not problem-specific (the \textit{init} and \textit{mixer} layers) are defined in generic terms, using qubit arrays. Note that variable domains that do not align with binary domains supported by type \lstinline{QNum} can be accommodated using problem constraints, in a manner discussed below. 

The focus of our example is the definition of the cost operator. The cost layer of the QAOA algorithm is an operator \(U(c,\ \gamma)\)  that applies a phase to computational-basis states in proportion to the problem-specific cost function \(c\) and the coefficient \(\gamma\). In our case, we wish to maximize the objective \(3a+5b\). We can directly apply the phase statement to rotate each state at an angle corresponding to the evaluation of this expression, factored by the coefficient parameter. We negate the expression to obtain a (negative) cost to be minimized.

Representing the objective of the problem is not enough. States that violate the maximum weight constraint \(2a+3b\le12\) need to be suppressed. In other words, we want to associate a high cost with violating states. A straightforward way to achieve this is to condition the application of the objective phase on the satisfaction of the constraint. We can compute the arithmetic (Boolean) condition onto a local variable and use it in a control statement. In this way, states that satisfy the constraint take on a negative phase proportionate to the objective, and states that violate the constraint are left unchanged. Using a \textit{within-apply} statement, we guarantee that the local variable is uncomputed and released. The implementation of the function \lstinline{cost_layer} is shown in \autoref{fig:knapsack_qaoa_cost_example}.

\begin{figure}[!htbp]
\begin{lstlisting}
@qfunc
def cost_layer(gamma: CReal, v: KnapsackVars):
  aux = QBit()
  within_apply(
    within=lambda: assign(2*v.a + 3*v.b <= 12, aux),
    apply=lambda: control(
      aux, lambda: phase(-(3*v.a + 5*v.b), gamma)
    ),
  )
\end{lstlisting}
\caption{
Code listing of the QAOA cost layer for an integer knapsack problem.
}
\label{fig:knapsack_qaoa_cost_example}
\end{figure}

It is often useful to reuse the same description of the objective and constraint in both the quantum subroutine and the classical logic exploring the parameter space, in order to evaluate them on the sampled results of the circuit execution. In Qmod/Python this can be done by simply factoring out these expressions to regular Python functions, since the same expression syntax is overloaded for classical and quantum interpretation with equivalent abstract semantics.

\section{Related work}

Python libraries such as Qiskit \cite{QiskitLibraryOnline} and PennyLane \cite{PennyLaneOnline} provide library implementations for quantum integer arithmetic building blocks, such as adders and multipliers. However, the construction of complex expressions must be carried out programmatically in a manner that does not reflect the high-level intent. Furthermore, a specific implementation must be chosen explicitly for each operation, further entangling the functional intent with implementation details.
The addition operation, for example, may be implemented in many ways \cite{draper2000, Cuccaro200, draper2004logarithmic, takahashi2009quantum, wang2016improved}.
Moreover, the uncomputation of intermediate results must be performed manually. This introduces an additional layer of complexity into the code even when done naively. The size of registers for intermediate results, the final result, and the number of auxiliary qubits for each operation must also be specified explicitly, further burdening the programmer.

Q\# is similar to Qiskit and PennyLane in providing a library of operations for basic arithmetic building blocks \cite{QsharpApiReference}. Unlike the Python libraries, Q\# supports local allocation and de-allocation of quantum storage. Hence, the management of auxiliary qubits can be handled within these building blocks and abstracted away in user code. Nevertheless, all implementation decisions within the basic operations, as well as in their composition into expressions, must still be handled explicitly. This leaves most of the responsibility for computing and uncomputing complex arithmetic to the programmer.

Qrisp \cite{Qrisp2024} is similar in spirit to Qmod, and specifically in its support for quantum numeric variables and native arithmetic expressions. In addition, Qrisp includes some unique features not found in Qmod, such as arbitrary modular arithmetic. However, there are important Qmod capabilities that Qrisp does not provide at the time of writing this paper. For instance, Qrisp requires the size and precision of numeric variables to be declared explicitly and does not support type inference in this regard. Therefore, unless the programmer makes strong assumptions about how expressions are implemented, they risk overestimating or underestimating the required size when declaring the left-hand side variable of an assignment. In addition, Qrisp neither automatically truncates least-significant bits nor provides control over operation precision bounds. Because precision is effectively unbounded, only integers are supported as classical multiplication operands.

While Qmod’s digital arithmetic (described in Section 3.1) has parallels in standard high-level programming languages, arithmetic computation in the phases or amplitudes of states (described in Section 3.2 and Section 3.3) does not. Of course, similar logic can be realized using gate-level description in any quantum programming framework. However, to the best of our knowledge, no other language currently supports these concepts natively, delegating their gate-level implementation to the compiler. In these respects, Qmod represents a significant step forward in terms of abstraction and automation potential.

\section{Conclusion}
In this paper, we presented Qmod, a high-level quantum programming language. Drawing on existing quantum programming concepts -- quantum variables, quantum functions, inversion and control functors, and conjugation operations -- Qmod also incorporates unique mechanisms and abstractions. Most notably, Qmod supports native expressions that combine quantum variables, classical variables, and literal constants using familiar arithmetic, relational, and bitwise operators. These expressions can be evaluated in multiple modes, encoding their results either digitally (in the computational basis) or non-digitally (in the phase or amplitude of basis states). Thus, Qmod lets programmers express complex numeric computations in high-level functional terms, leaving implementation details to automation. Challenging implementation tasks, such as representing intermediate and final results compactly, uncomputing intermediate results efficiently, and mapping to gate-level descriptions, are taken care of by the compiler.

These capabilities address a significant bottleneck in the development of quantum applications. Arithmetic computations are an important aspect of many quantum algorithms, in themselves and as means to efficiently approximate non-arithmetic functions. Moreover, arithmetic expressions naturally capture classical problem-domain logic, such as search predicates, cost functions, and payoff functions. Offloading the circuit-level synthesis of these descriptions to robust automation tools dramatically improves the quality and scalability of the resulting circuits \cite{classiq2024}, and boosts the productivity of algorithm and application developers.

In this paper, we focus primarily on Qmod's language constructs, semantics, and usage. A detailed description of the algorithms for synthesizing these constructs into optimal gate-level descriptions is left for future work. Further investigations will include quantitative evaluation of Qmod in a broader set of use cases, comparing it with other languages and frameworks. This comparison should examine the impact of adaptive numeric storage and precision decisions, as well as other implementation factors, on resource requirements and the quality of results for large-scale applications.

\section*{Platform and code availability}
All code examples in this paper were tested using the Classiq platform -- a comprehensive compiler and development environment for Qmod. Instructions on how to access the Python SDK, as well as a complete language reference manual, are found at \href{https://docs.classiq.io}{docs.classiq.io}. A broad collection of application examples is hosted at \href{https://github.com/Classiq/classiq-library}{github.com/Classiq/classiq-library}.

\bibliographystyle{IEEEtran} 
\bibliography{qmod_paper}  
\end{document}